\documentclass[epj]{svjour}

\usepackage[autostyle,italian=guillemets]{csquotes}
\usepackage[utf8]{inputenc}
\usepackage[nowrite,infront,standard]{frontespizio}
\usepackage[english]{babel}
\usepackage{physics}
\usepackage{amsmath,amssymb}
\usepackage [svgnames]{xcolor}
\usepackage{mathrsfs}
\usepackage{subfloat}
\usepackage{frontespizio}
\usepackage{chemmacros}
\usepackage{float}
\usepackage{graphicx} 
\usepackage{bm}
\usepackage{appendix}
\usepackage{hyperref}
\usepackage{midpage}
\usepackage{placeins}
\usepackage{subfig}

\begin{document}
\title{Wave packet approach to quantum correlations in neutrino oscillations}
\author{Massimo Blasone\inst{1,2}, Silvio De Siena\inst{3} and Cristina Matrella\inst{1,2} 
}                     
%
%
\institute{ Dipartimento di Fisica, Universit\`a degli Studi di Salerno, Via Giovanni Paolo II, 132 84084 Fisciano, Italy \and INFN, Sezione di Napoli, Gruppo Collegato di Salerno, Italy \and Retired Professor, Universit\`a degli Studi di Salerno,  email: silvio.desiena@gmail.com}
\date{Received: date / Revised version: date}
%
\abstract{
Quantum correlations provide a fertile testing ground for investigating fundamental aspects of quantum physics in various systems, especially in the case of  relativistic (elementary) particle  systems as neutrinos.
In a recent paper, Ming et al. (Ming et al. Eur.~Phys.~J.~C 80 (2020) 275), in connection with results of Daya-Bay and MINOS experiments, have studied the quantumness in neutrino oscillations in the framework of plane-wave approximation. We extend their treatment by adopting  the wave packet approach that  accounts for effects due to localization and decoherence.  This leads to a better agreement with experimental results, in particular for the case of MINOS experiment.
\PACS{
      {PACS-key}{discribing text of that key}   \and
      {PACS-key}{discribing text of that key}
     } 
} 
\authorrunning{M.~Blasone, S.~De~Siena and C.~Matrella}
\maketitle
\section{Introduction}
\label{intro}

The study of quantum correlations \cite{Adesso} is a very active research area in view of  applications such as quantum communication and computation, and quantum
cryptography. They have been studied in a variety of physical contexts, such as quantum optics
and condensed matter systems but, more recently, attention has also been directed towards subatomic
physics. A particular focus has been
concentrated on relativistic systems of neutrinos and mesons \cite{Dixit}-\cite{Song}, which are interesting candidates for applications of quantum information beyond photons; investigations in this direction can also provide a possible ``feedback'' effect allowing better understanding of  fundamental physical properties of such particles.

The phenomenon of  neutrino oscillations offers a rare example of quantum correlations on macroscopic scale.
Neutrino oscillations have been investigated both from a theoretical perspective, and in relation to the
available data from several experiments, confirming the intrinsic quantum nature of this phenomenon \cite{formaggio}. One
of the most important and useful aspects concerning quantum correlations in neutrinos is that they can be
expressed in terms of the oscillation probabilities, which are directly obtainable from experiments.

In a recent article \cite{Ming}, Ming et al. have investigated quantum correlations in neutrino oscillations by referring to  Daya Bay  \cite{DAYA,DBWP}, and MINOS experiments \cite{MINOS1,MINOS2}. They found
interesting results by investigating the violation of classical bounds by quantum markers such as the nonlocal advantage of quantum coherence
(NAQC) and the Bell nonlocality, which detect different levels of quantumness.

Their results have been obtained in the framework of the plane-wave approach which, as well known, does not account for  the effects due to
localization and decoherence. Thus, a more realistic description of this phenomenon requires the
wave-packet approach for neutrino oscillations  introduced in Refs.\cite{Giunti,Giunti2}.

In this paper,  adopting the wave-packet approach, we  study quantum correlations associated to neutrino oscillations, highlighting that localization and decoherence effects induce  attenuation and limitations in the spatial extension of the correlations. By explicitly referring to the case of Daya Bay \cite{DAYA} and MINOS experiments,  we compare our results with those of Ming et al. \cite{Ming}. Our results are generally different from those of Ref.\cite{Ming}: however, in the case of Daya Bay, the effect of corrections due to wave packet approach is negliglible, as already remarked in Ref.\cite{DBWP}, while for the case of MINOS experiment, the corrections are very relevant since they lead to a much improved fit of experimental data.

The plan of the paper is as follow: In Section 2 we  recall the notions of NAQC and Bell nonlocality.
In Section 3 we review the study of Ming et al. carried out in the plane-wave approximation. 
In Section 4, we generalize  the study of Section 3 within the
wave-packet approach to neutrino oscillations and compare our results with those obtained in Ref.\cite{Ming}. Section 5 is devoted to conclusions and outlook.
Some appendices containing some technical issues are also provided.

\section{NAQC and Bell nonlocality}
\label{sec:0}
In this section we briefly review some definitions and properties of NAQC and Bell nonlocality, following Refs. \cite{Hu},\cite{Mondal}.\\
 A state is said to be coherent provided that there are nonzero elements in the non diagonal position of its density matrix representation. There are various ways to quantify the coherence of a state. One of these is $l_{1}$-norm of coherence, which is given by:

\begin{equation}
C_{l_{1}}(\rho)=\sum_{i\ne j}|\rho_{i,j}|
\label{0.1}
\end{equation}

 Quantum coherence can also be linked to quantum correlations, although they are defined in different scenarios and capture different aspects of the quantumness of a state.\\
We study the effect of non locality on quantum coherence in a bipartite scenario, so that it can be applied to the case of two-flavor neutrino oscillations. \\
Let us consider the $l_{1}$-norm of coherence of a state $\rho$. If a qubit is prepared in either spin up or spin down state along z-direction then the qubit is incoherent when we calculate the coherence in z-basis $(C_{z}^{l_{1}}=0)$ and it is fully coherent in x- and y-basis $(C_{x(y)}^{l_{1}}=1)$ . One may ask what is the upper bound of $C^{l_{1}}=C_{x}^{l_{1}} + C_{y}^{l_{1}} + C_{z}^{l_{1}}$. This limit for a general qubit state $\rho$ is given by:

\begin{equation}
\sum_{i=x,y,z}C_{i}^{l_{1}}(\rho)\le C_{max},
\label{0.2}
\end{equation}
where  $C_{max}=\sqrt{6}$ is the state-independent upper bound. The equality sign holds for a pure state $\rho_{max}=\frac{1}{2}\bigl[\frac{1}{\sqrt{3}}(\sigma_{x}+\sigma_{y}+\sigma_{z}\bigl)]$.
 A violation of this inequality by the conditional states of a part of the system implies that one can achieve a non-local advantage of quantum coherence.\\
Now we introduce another criterion for NAQC via the steering game \cite{Ming}. Let us suppose that Alice and Bob are two game participants and share qubits A and B with state $\rho_{AB}$, respectively. Alice performs a measurement $\Pi_{i}^{b}$ on A and obtains the outcome $b=\{0,1\}$ with probability $p_{\Pi_{i}^{b}}$.
The measured state for the two-qubit state can be obtained as $\rho_{AB|\Pi_{i}^{b}}=(\Pi_{i}^{b}\otimes I)\rho_{AB}(\Pi_{i}^{b}\otimes I)/p_{\Pi_{i}^{b}}$ and the conditional state for qubit B is $\rho_{B|\Pi_{i}^{b}}=Tr_{A}(\rho_{AB|\Pi_{i}^{b}})$. Then Alice tells Bob her measurement choice and Bob has to measure the coherence of qubit B at random in the eigenbases of the other two Pauli matrices $\sigma_{j}$ and $\sigma_{k}$.\\
 A violation of (\ref{0.2}) by the conditional states of a part of the system implies that one can achieve a non-local advantage of quantum coherence. The criterion for achieving a NAQC of qubit B can be written as:

\begin{equation}
N^{l_{1}}(\rho_{AB})=\frac{1}{2}\sum_{i,j,b}p(\rho_{\Pi_{j\ne i}}^{b})C_{l_{1}}^{\sigma_{i}}(\rho_{B|\Pi_{j\ne i}})> \sqrt{6}.
\label{0.3}
\end{equation}

In  Ref. \cite{Ming} it is shown that NAQC is  a stronger quantum correlation than Bell nonlocality. This latter can be detected by the violation of the Clauser-Horne-Shimony-Holt (CHSH) inequality $B(\rho_{AB})=|\langle B_{CHSH}\rangle|\le2$.
 If this inequality is violated then the states are Bell nonlocal. It means that the classical theories cannot describe the system of interest.
The Bell-CHSH inequality can be also written as:
\begin{equation}
M(\rho_{AB})=\max(u_{i}+u_{j})\le1, \hspace{1.5cm} i\ne j.
\label{0.4}
\end{equation}

 Here, $\rho_{AB}$ is the density matrix associated with the state of interest. $u_{i}\hspace{0.1cm} (i=1,2,3)$ are the eigenvalues of the matrix $T^{\dagger}T$, where $T_{m,n}=\Tr[\rho(\sigma_{m}\otimes \sigma_{n})]$ are the elements of a correlation matrix $T$.

\section{Quantum correlations in neutrino oscillations -- plane waves}
\label{sec:1}
Following Ref. \cite{Ming} we now study quantum correlations in the plane wave approach in 
 the case of two-flavor oscillations.\\
The time evolution of the state for two-flavor neutrino oscillations gives us:

\begin{equation}
\ket{\nu_{\alpha}(t)}=a_{\alpha\alpha}(t)\ket{\nu_{\alpha}}+a_{\alpha\beta}(t)\ket{\nu_{\beta}},
\label{1.1}
\end{equation}
with $\alpha,\beta=e,\mu$.
From this equation is simple to see that the survival probability to find a neutrino of flavor $\alpha$ after a time $t$ is given by $P_{\alpha\alpha}(t)=|a_{\alpha\alpha}(t)|^{2}$, while the transition probability is given by $P_{\alpha\beta}(t)=|a_{\alpha\beta}(t)|^{2}$.
First we see that it is possible to rewrite Eqs.(\ref{0.3}) and (\ref{0.4}) for the NAQC and the Bell nonlocality in terms of neutrino oscillation probabilities (see Appendix \ref{Appendice A}) as:

\begin{equation}
N^{l_{1}}(\rho_{AB})=2+2\sqrt{P_{\alpha\alpha}(t)(1-P_{\alpha\alpha}(t))}>2.45,
\label{1.2}
\end{equation}
and
\begin{equation}
M(\rho_{AB})=1+4 P_{\alpha\alpha}(t)(1-P_{\alpha\alpha}(t))\le1.
\label{1.3}
\end{equation}

\vspace{0.5cm}
The survival probability is given by:

\begin{equation}
P_{\alpha\alpha}(L)=1-\sin^{2}{2\theta}\sin^{2}\biggl(\frac{\Delta m^{2}}{4c\hbar}\frac{x}{E}\biggl)
\label{1.4}
\end{equation}
where $\theta$ is the mixing angle, $\Delta m^{2}$ is the mass-squared difference, $E$ is the neutrino energy and $L=ct$ is the distance between the production and the detection points after a time $t$.

\begin{figure}[t]
\centering
 \subfloat[]{{\includegraphics[width =7 cm]{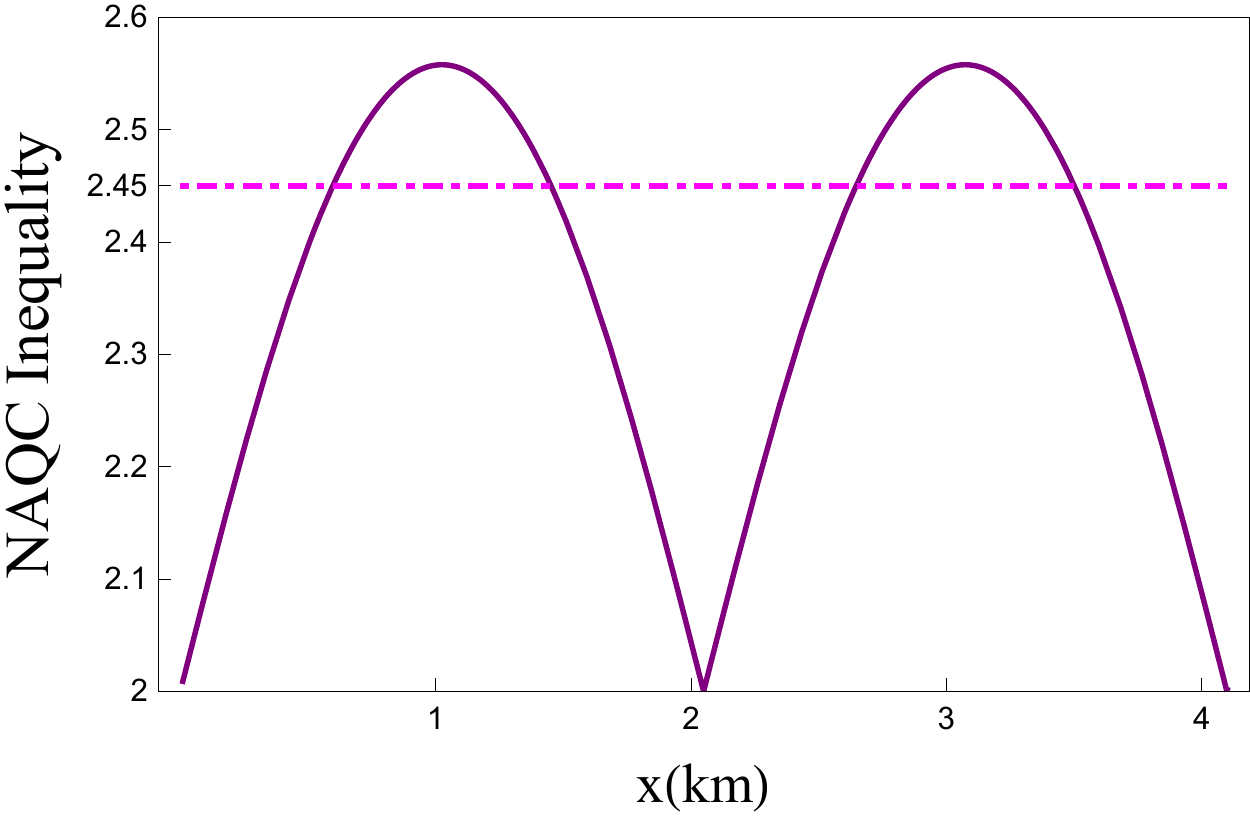}}\quad\quad
{\includegraphics[width =7 cm]{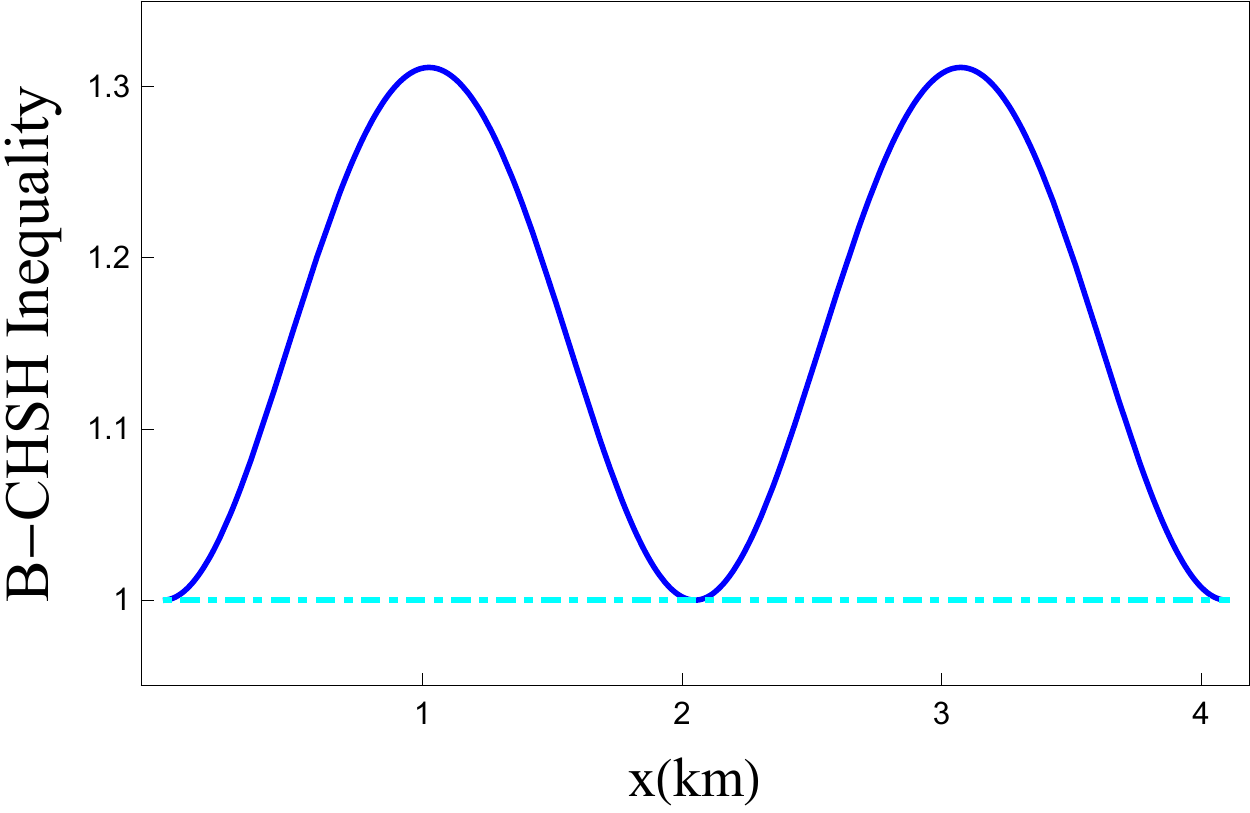}}}\\
 \subfloat[]{{\includegraphics[width =7 cm]{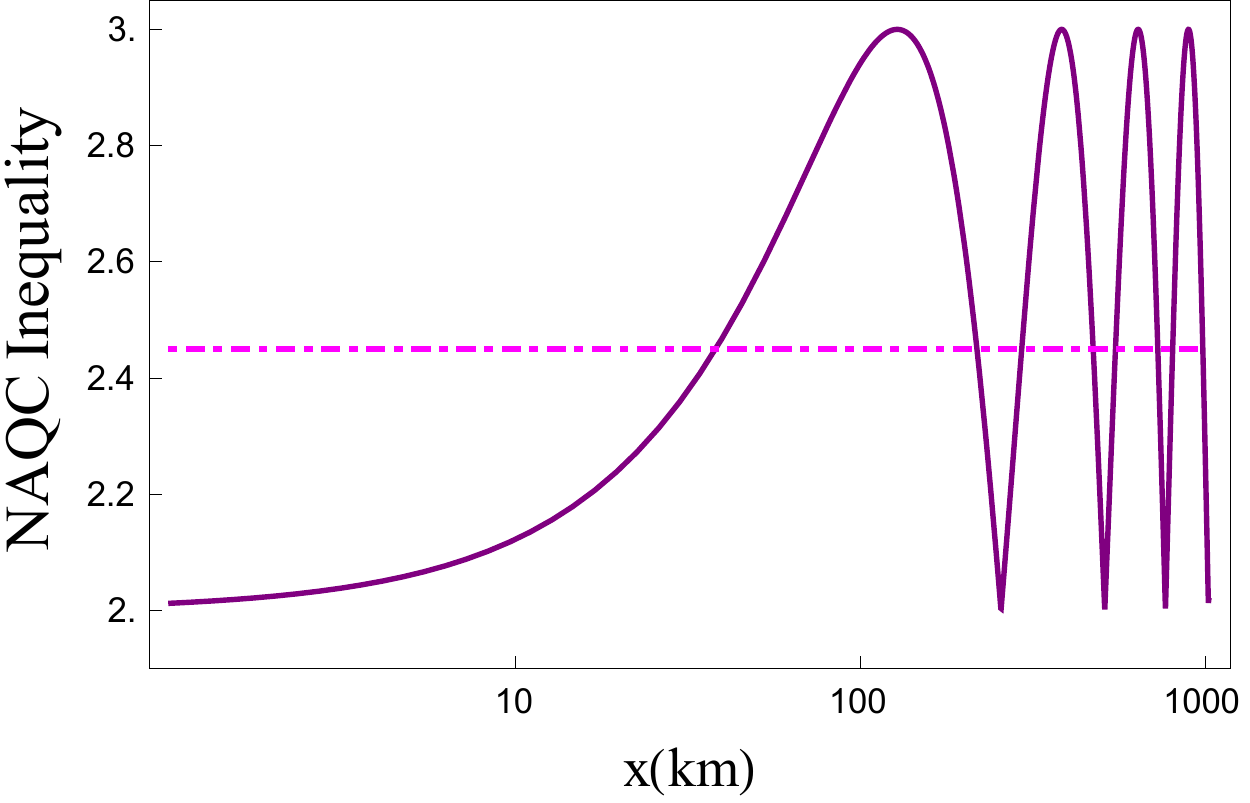}}\quad\quad
{\includegraphics[width =7 cm]{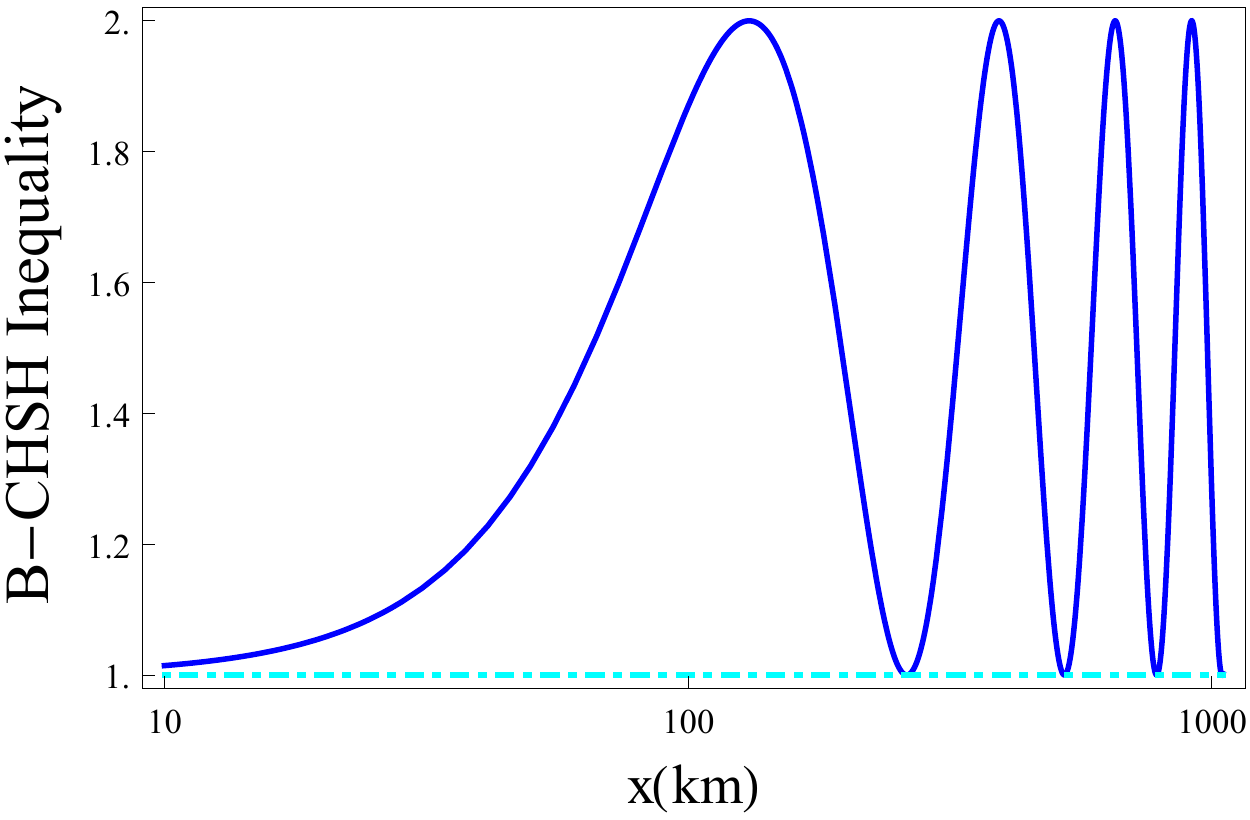}}}
\caption{ NAQC and Bell-CHSH inequalities as a function of the distance. 
(a) The plot is made using the data from Daya Bay experiment: $\sin^{2}2\theta_{13}=0.084^{+0.005}_{-0.005}$ and $\Delta m_{ee}^{2}=2.42_{-0.11}^{+0.10}\times10^{-3} eV^{2}$. The value of the energy is $E=2 MeV$. 
(b)The plot is made using the data from MINOS experiment: $\sin^{2}2\theta_{23}=0.95^{+0.035}_{-0.036}$ and $\Delta m_{32}^{2}=2.32_{-0.08}^{+0.12}\times10^{-3} eV^{2}$. The value of the energy is $E=0.5 GeV$. The $x$-axis is in logarithmic scale. 
The magenta and cyan dot-dashed horizontal lines are the bounds of the NAQC and Bell-CHSH inequalities, respectively.  }
\label{Fig:01}
\end{figure}

In Fig. [\ref{Fig:01}] we show the violations of the NAQC and Bell-CHSH inequalities, using the data from the Daya Bay Reactor Neutrino
 \cite{DAYA,DBWP} and MINOS\cite{MINOS1,MINOS2} experiments, as reported in Ref.\cite{Ming}. \\

Note that, while in Ref. \cite{Ming}, the inequalities are plotted as a function of $L/E$, here we express them as a function of distance $x$ alone. This will be useful for making the comparison with the wave packet treatment of next section.

In Fig. [\ref{Fig:01}] a violation of these inequalities means a strong quantumness. On the left panels we see how we can reach a non local advantage of quantum coherence for certain regular range of distances, strongly dependent on the oscillation probability of the neutrino. On the other hand, on the right panels we observe that the Bell nonlocality is present for all values of the distance $x$. 
As highlighted in Ref. \cite{Ming}, this means that NAQC is a stronger quantum correlation than Bell nonlocality.

\section{Quantum correlations in neutrino oscillations -- wave packets}
\label{sec:2}
 In this section, we use the wave packet approach to neutrino oscillations to extend the result of the previous section.

In this approach, the (\ref{1.1}) becomes:

\begin{equation}
|\nu_{\alpha}(x,t)\rangle=\sum_{j}U^{*}_{\alpha j}\psi_{j}(x,t)|\nu_{j}\rangle,
\label{2.1}
\end{equation}
where $U_{\alpha j}$ denotes the elements of the PMNS mixing matrix. $\psi_{j}(x,t)$ is the wave function of the mass eigenstate $|\nu_{j}\rangle$ with mass $m_{j}$:

\begin{equation}
\psi_{j}(x,t)=\frac{1}{\sqrt{2\pi}}(2\pi\sigma_{p}^{P^{2}})^{-\frac{1}{4}}\int dp \hspace{0.1cm}\exp\biggl\{-\frac{(p-p_{j})^{2}}{4\sigma_{p}^{P^{2}}}\biggl\}e^{ipx-iE_{j}(p)t}, 
\label{2.2}
\end{equation}
where we assume a Gaussian distribution for the momentum of the massive neutrino $\nu_{j}$. From (\ref{2.2}) it is possible to obtain the neutrino oscillation probability in the wave packet approach  (see Appendix \ref{Appendice B}).\\

\subsection{Electron neutrino oscillations}
\begin{figure}[h]
\centering
{\includegraphics[width =7 cm]{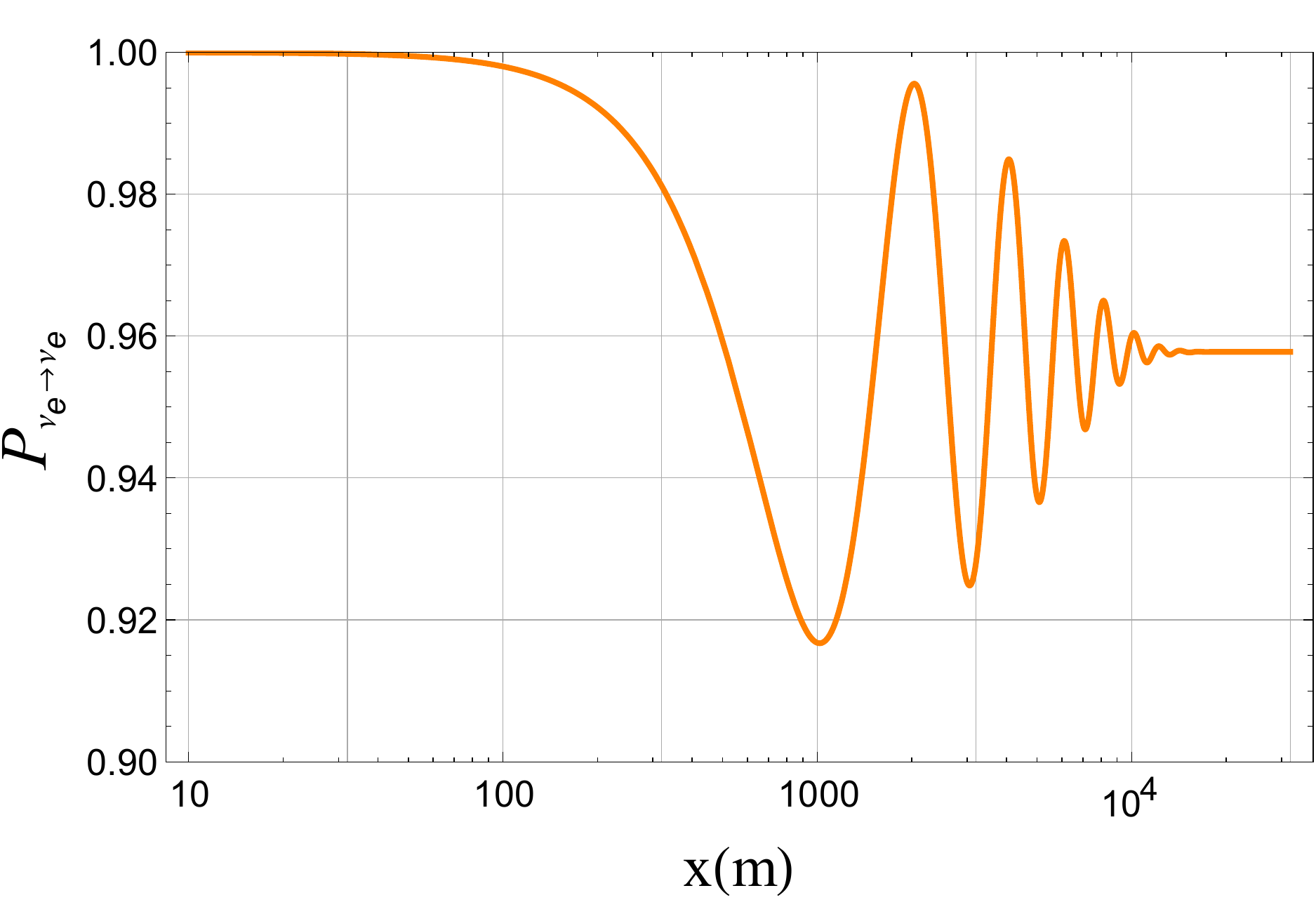}}\quad\quad
{\includegraphics[width =7 cm]{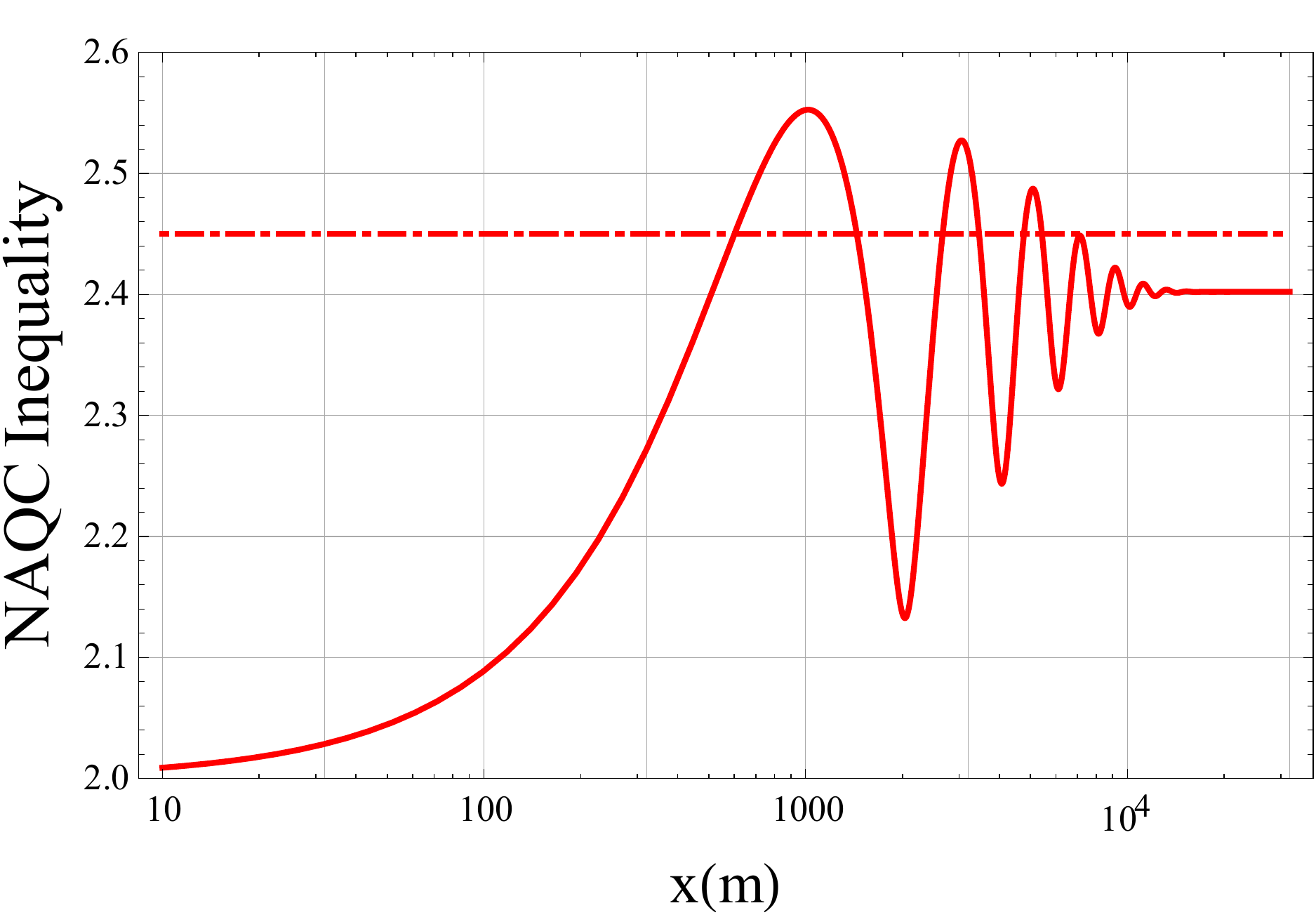}}
\caption{On the left panel is shown the survival transition for an electronic neutrino in the wave packet approach. The plot is done with the following values of parameters: $E=2\, MeV$, $\xi=0$,  $\sin^{2}2\theta_{13}=0.084\pm0.005$ and $\Delta m_{ee}^{2}=2.42_{-0.11}^{+0.10}\times10^{-3} eV^{2}$ and $\sigma_{x}=3.3\times10^{-6} m$. The x-axis is in logarithmic scale. On the right panel is shown the NAQC inequalities for this survival probability. The  dot-dashed horizontal line is the bound of the NAQC inequality. }
\label{Fig:01.1}
\end{figure}

In order to compare the plane waves and the wave packet approaches to neutrino oscillation, we start considering an electronic neutrino at the initial time $t=0$.
In fig.[\ref{Fig:01.1}] is plotted the formula (\ref{sub10}) for the electronic neutrino survival probability together with the NAQC inequality as  functions of the distance, in the wave packet approach. 


\begin{figure}[h]
\centering
{\includegraphics[width =7 cm]{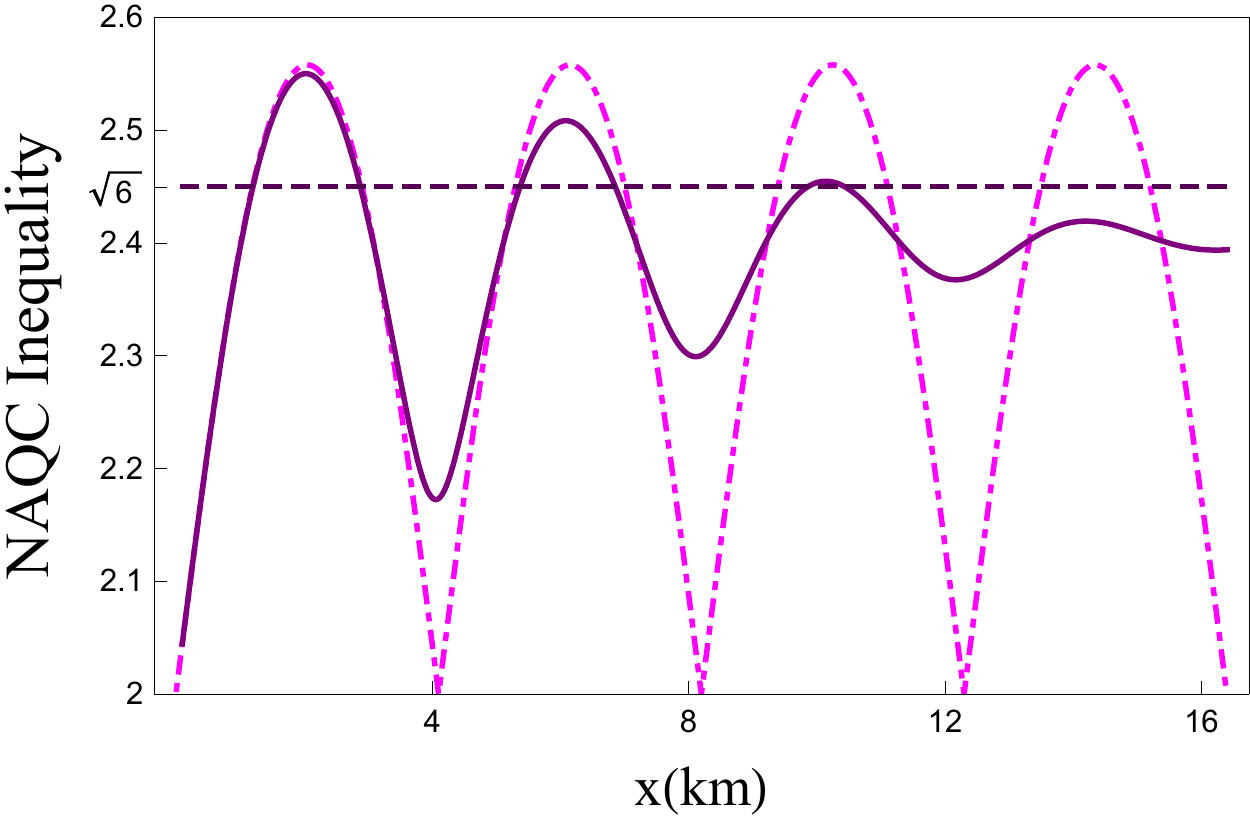}}\quad\quad
{\includegraphics[width =7 cm]{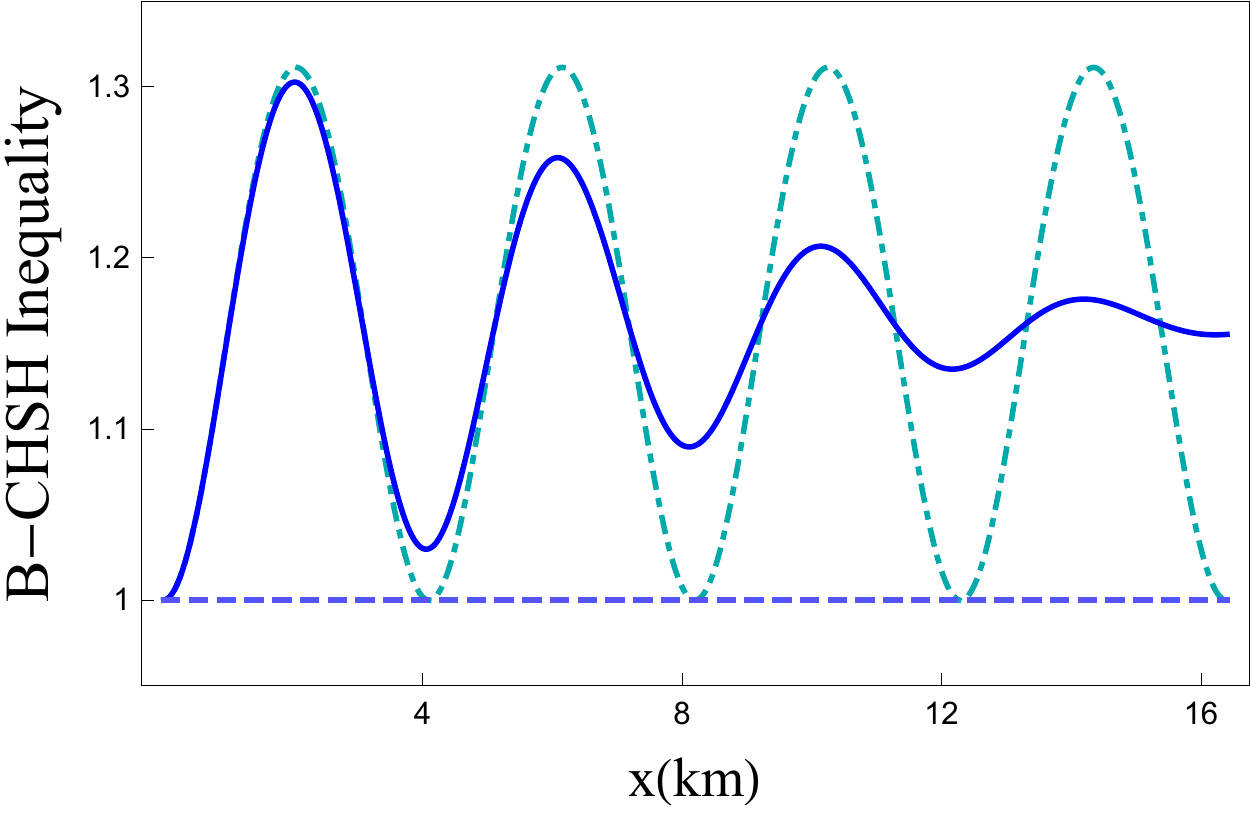}}
\caption{NAQC and Bell-CHSH inequalities as a function of the distance. The plot is made using the data from Daya Bay experiment: $\sin^{2}2\theta_{13}=0.084\pm0.005$ and $\Delta m_{ee}^{2}=2.42_{-0.11}^{+0.10}\times10^{-3} eV^{2}$  and $\sigma_{x}=1.25\times10^{-6} m$. The value of the energy is $E=4 MeV$. The darker magenta and the lighter blue dashed horizontal lines are the bounds of the NAQC and Bell-CHSH inequalities, respectively. The solid and dot-dashed  lines represent the plot for the wave packet approach and plane waves approximation, respectively. }
\label{Fig:02}
\end{figure}

In Fig.[\ref{Fig:02}] we compare the plots of the NAQC and the Bell-CHSH inequalities obtained with the approximation of plane waves and those obtained with the wave packet approach. On the right panel of the figure we observe a violation of the Bell inequality for each value of the distance $x$. Nevertheless, from a certain distance onwards the violation decreases until it reaches a constant value for large $x$. Certainly the most interesting behavior is observed on the left panel of the figure. We can see how we can still reach a non local advantage of quantum coherence, but only up to a certain distance. Indeed at great distances we go down the value $\sqrt{6}$ due to the spatial separation of the wave packets. The effects of interference are destroyed by the decoherence due to localization.\\


\begin{figure}[h!]
\centering
\includegraphics[width =9 cm]{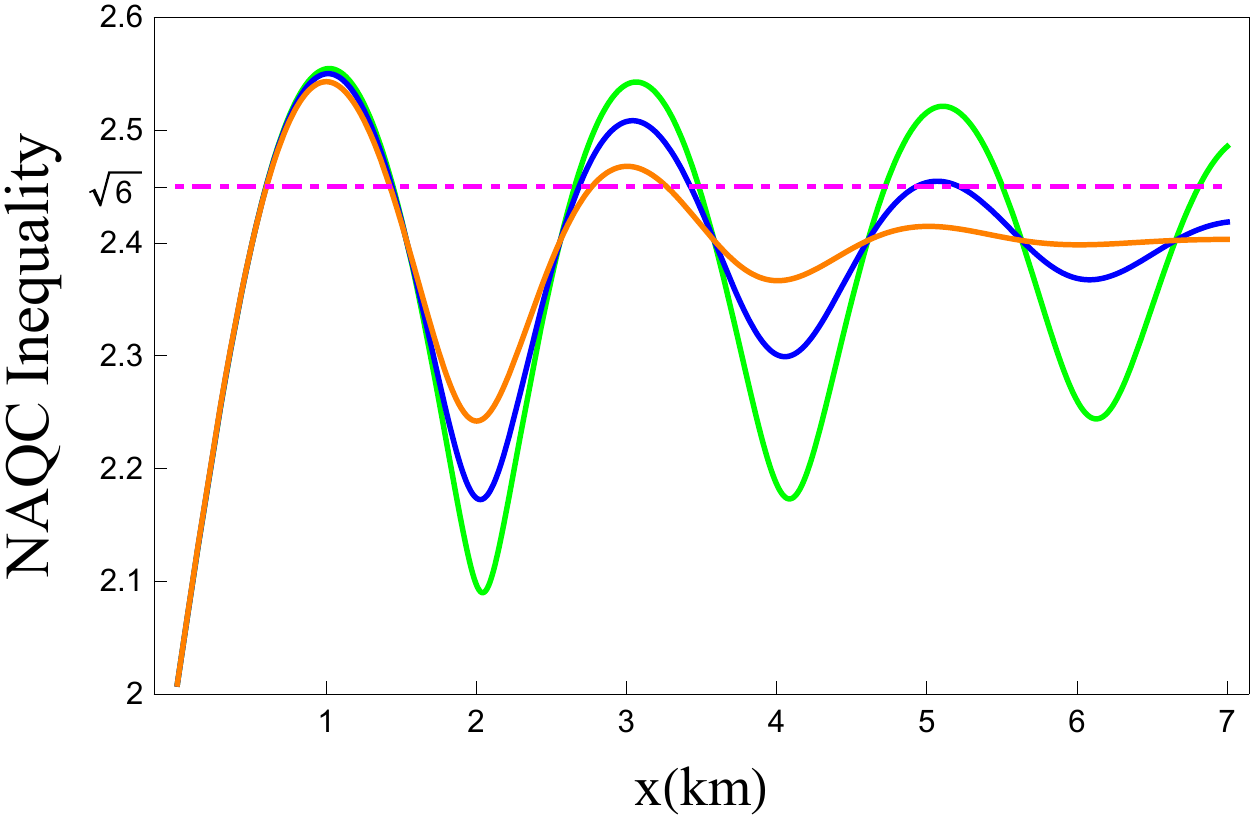}
\caption{NAQC inequality as a function of the distance  for three different wave packet widths $\sigma_{x}$: $\sigma_{x}=5\times10^{-6}$ (green line),  $\sigma_{x}=2.5\times10^{-6} m$ (blue line) and  $\sigma_{x}=1.7\times10^{-6} m$ (orange line). The value of the energy is $E=2 MeV$.The  dot-dashed horizontal line is the bound of the NAQC inequality.  }
\label{Fig:04}
\end{figure}

Another interesting behavior that emerges from the wave packet treatment is that the amount of coherence depends by the wave packet width $\sigma_{x}$. In Fig.[\ref{Fig:04}] is shown as it increases by $\sigma_{x}$. This behavior is due to the overlapping of the mass eigenstates that increases by $\sigma_{x}$ and more coherence is expected \cite{Ettefaghi}.


\newpage
\subsection{Muon neutrino oscillations}

\begin{figure}[h]
\centering
{\includegraphics[width =7 cm]{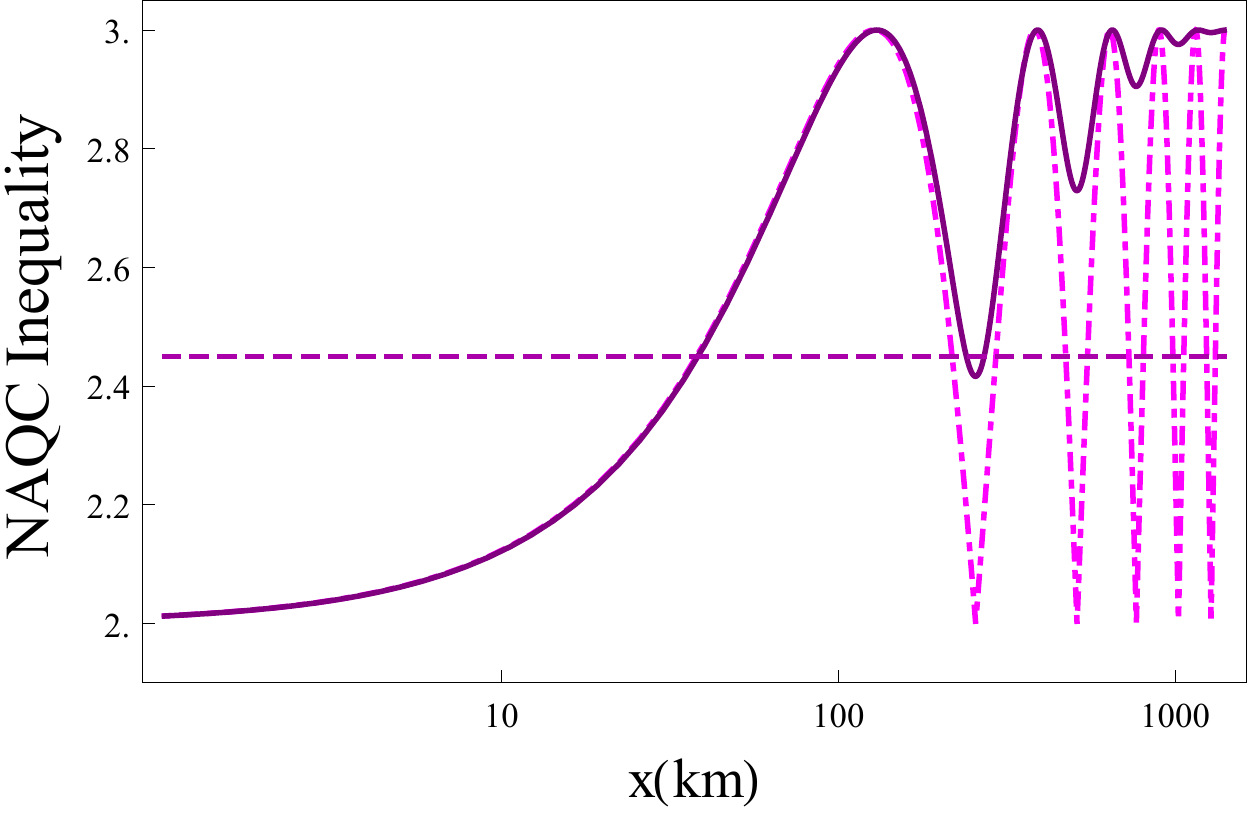}}\quad\quad
{\includegraphics[width =7 cm]{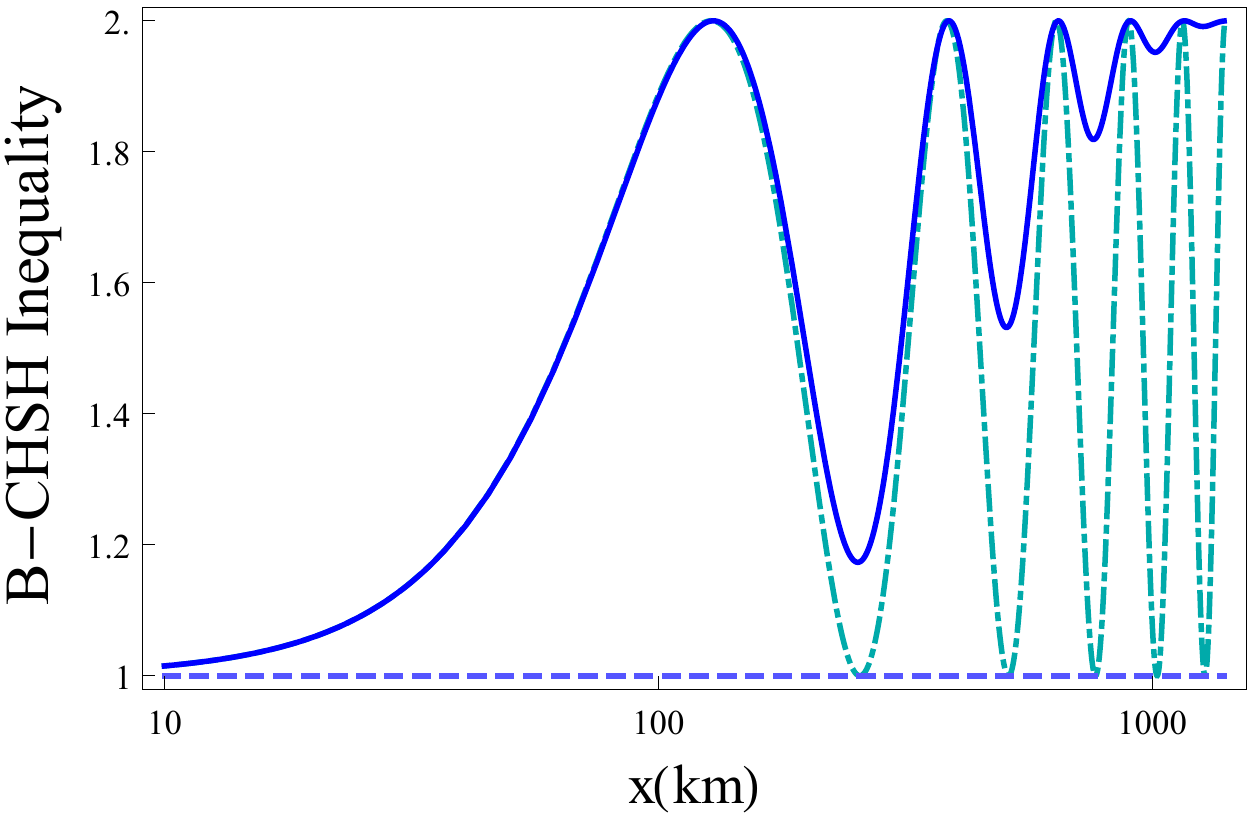}}
\caption{NAQC and Bell-CHSH inequalities as a function of the distance.  The plot is made using the data from MINOS experiment: $\sin^{2}2\theta_{23}=0.95^{+0.035}_{-0.036}$ and $\Delta m_{32}^{2}=2.32_{-0.08}^{+0.12}\times10^{-3} eV^{2}$. The value of the energy is $E=0.5 GeV$ and  $\sigma_{x}=7\times10^{-9} m$. The $x$-axis is in logarithmic scale.  The darker magenta and the lighter blue dashed horizontal lines are the bounds of the NAQC and Bell-CHSH inequalities, respectively. The solid and dot-dashed  lines represent the plot for the wave packet approach and plane wave  approximation, respectively. }
\label{Fig:05}
\end{figure}

Now, we consider the case of MINOS experiment, which deals with a muon neutrino at the initial time. In this case, the length and energy scales involved are very different from the case of Daya-Bay experiment. 
In Fig.[\ref{Fig:05}], using the same parameter values as in Ref.\cite{Ming}, we compare the plots of the NAQC and the Bell-CHSH inequalities obtained with the approximation of plane waves and those obtained with the wave packet approach.

It is evident from  Fig.[\ref{Fig:05}] that exists a considerable difference between the two approaches. On the left panel, we see that we reach a non local advantage of quantum coherence for any value above some distance, which does not occur for the plane wave approach. From Fig.(5) of Ref.\cite{Ming}, where also experimental points are shown, it is clear that the present approach based on wave packets fits experimental data considerably better than the plane wave curve.

For the case of Bell nonlocality, both approaches give curves above the bound, but again the fit by wave packet curve appears to be sensibly better due to the attenuation of the oscillations on the distance scale involved.

In definitive, our results show how in the case in which  long spatial extensions and high energies are involved, the wave packet approach turns out to be fundamental for a more realistic description of neutrino oscillations.

\newpage
\section{Conclusions}

In this paper we have extended a recent study by Ming et al.\cite{Ming}, by adopting a  more realistic, wave-packet approach, in contrast with their treatment based on plane waves. In particular, we have considered two quantificators of quantumness there studied, namely the nonlocal advantage of quantum coherence (NAQC) and Bell localization, which in the wave-packet approach, exhibit a non-trivial dependence on distance and energy.\\
It is to be pointed out that in the literature there exists a debate on the necessity of adopting wave packet approach with respect to the plane wave approximation. In this paper we show that, although in some experimental situations plane wave approach is sufficient, this is not true in other experiment characterized by very different parameter values. 

Infact we found that, in the case of Daya Bay experiment, the wave packet treatment does not add significant corrections to the result by Ming et al.\cite{Ming}. This is in agreement with the analysis of Ref.\cite{DBWP}, where it was shown that plane waves are sufficient to describe rather accurately such short-baseline, low energy, neutrino oscillation experiment.

On the other hand, in the MINOS experiment, due to the long baseline and high energies involved, we found a remarkable correction, and a much better fit of experimental data, of our treatment with respect to the plane wave analysis of Ref.\cite{Ming}. In particular, our fit accounts for a NAQC marker even beyond the bound and both in the NAQC and Bell nonlocality cases, shows the attenuation phenomenon along the length scale of the experiment.This is due to a longer spatial extension and a greater energy  of the MINOS with respect to the Daya Bay experiment.

We would like to remark one important aspect concerning the neutrino wave packet dispersion $\sigma_x$, whose value is not apriori known, as also discussed in  Ref.\cite{DBWP}.  There a wide range of values for such parameter was indicated, which allows us to agree reasonably well with the experimental values for the quantum markers reported in Ref.\cite{Ming}.

We plan to extend our study to the case of  three-flavor neutrino oscillations, which could be interesting from a theoretical point of view,  due to the presence of the CP violation phase. Furthermore,  a similar approach can be exploited for studying correlations of other particles, as  mesons, also taking into account  other quantum markers, beyond those here exploited.

Finally, we plan to consider the  extension of present work in the framework of the quantum field theory approach to neutrino mixing and oscillations \cite{Blasone:1995aop,Blasone:1998hf}. In particular, in Ref.\cite{spaceoscill}, neutrino oscillations  have been studied by means of wave packets and relativistic flavor currents, which give a complete characterization of the space-time features of this phenomenon and which should then account for the quantum correlations considered in this paper.

\newpage

\section*{Appendix A: NAQC and Bell nonlocality criterions in terms of neutrino oscillation probabilities.}
\label{Appendice A}
We consider the state:

\begin{equation}
\ket{\nu_{\alpha}(t)}=a_{\alpha\alpha}(t)\ket{\nu_{\alpha}}+a_{\alpha\beta}(t)\ket{\nu_{\beta}},
\label{a1}
\end{equation}
 with $\alpha,\beta=e,\mu$.\\
The corresponding density matrix is given by:

\begin{equation}
\rho^{\alpha}_{AB}(t)=
\begin{pmatrix}
0&0&0&0\\
0&|a_{\alpha\beta}(t)|^{2}&a_{\alpha\beta}(t)a_{\alpha\alpha}^{*}(t)&0\\
0&a_{\alpha\alpha}(t)a_{\alpha\beta}^{*}(t)&|a_{\alpha\alpha}(t)|^{2}&0\\
0&0&0&0
\end{pmatrix}
\label{a2}
\end{equation}
in the orthonormal basis $\{\ket{00}, \ket{01},\ket{10},\ket{11}\}$.

\subsection*{A.1: NAQC}
We first see how we can write the NAQC criterion in terms of neutrino oscillation probability.
 We want to perform Pauli measurement $\sigma_{x}$ on qubit $A$. For this aim, the post measurement states for the initial electron flavor state $\rho^{\alpha}_{AB}$ are expressed as:

\begin{equation}
\rho^{\alpha}_{AB|\sigma_{x_{k}}}(t)=[(\ket{x_{k}}\bra{x_{k}}\otimes \textbf{1})\rho^{\alpha}_{AB}(t)(\ket{x_{k}}\bra{x_{k}}\otimes \textbf{1})]/p_{\sigma_{x_{k}}},
\label{a3}
\end{equation}
where:

\begin{equation}
p_{\sigma_{x_{k}}}=\Tr[(\ket{x_{k}}\bra{x_{k}}\otimes \textbf{1})\rho^{\alpha}_{AB}(t)(\ket{x_{k}}\bra{x_{k}}\otimes \textbf{1})]
\label{a4}
\end{equation}

Here $\ket{x_{k}} (k=1,2)$ are the eigenstates of Pauli observables $\sigma_{x}$.

The conditional state for particle $B$ can be expressed as:
\begin{equation}
\rho_{B|\sigma_{x_{k}}}=\Tr_{A}\bigl(\rho^{\alpha}_{AB|\sigma_{x_{k}}}(t)\bigl).
\label{a5}
\end{equation}
The $l_{1}$-norm coherence for the conditional state for $B$ in the basis of eigenvector of Pauli observales $\sigma_{y}$ and $\sigma_{z}$  can be obtained as:

\begin{equation}
C^{\sigma_{x_{k}}}_{l_{1}}\bigl(\rho_{B|\sigma_{x_{k}}}\bigl)=\bigl|\langle y_{1}|\rho_{B|\sigma_{x_{k}}}|y_{2}\rangle\bigl|+ \bigl|\langle y_{2}|\rho_{B|\sigma_{x_{k}}}|y_{1}\rangle\bigl|
+\bigl|\langle z_{1}|\rho_{B|\sigma_{x_{k}}}|z_{2}\rangle\bigl|+\bigl|\langle z_{2}|\rho_{B|\sigma_{x_{k}}}|z_{1}\rangle\bigl|.
\label{a6}
\end{equation}
We show the explicit calculation for $C^{\sigma_{x_{1}}}_{l_{1}}\bigl(\rho_{B|\sigma_{x_{1}}}\bigl)$.

We remember that:

\begin{equation}
\ket{x_{1}}=\frac{1}{\sqrt{2}}
\begin{pmatrix}
1\\
1
\end{pmatrix}
\label{a7}
\end{equation}

Then, we have:
\begin{equation}
(\ket{x_{1}}\bra{x_{1}}\otimes \textbf{1})=
\begin{pmatrix}
1&1\\
1&1
\end{pmatrix}
\otimes
\begin{pmatrix}
1&0\\
0&1
\end{pmatrix}
=
\frac{1}{2}
\begin{pmatrix}
1&0&1&0\\
0&1&0&1\\
1&0&1&0\\
0&1&0&1
\end{pmatrix}
\label{a8}
\end{equation}

\begin{equation}
\rho^{\alpha}_{AB|\sigma_{x_{1}}}(t)=
\begin{pmatrix}
0&0&0&0\\
a_{\alpha\beta}(t)a_{\alpha\alpha}^{*}(t)&|a_{\alpha\beta}(t)|^{2}&a_{\alpha\beta}(t)a_{\alpha\alpha}^{*}(t)&|a_{\alpha\beta}(t)|^{2}\\
|a_{\alpha\alpha}(t)|^{2}&a_{\alpha\alpha}(t)a_{\alpha\beta}^{*}(t)&|a_{\alpha\alpha}(t)|^{2}&a_{\alpha\alpha}(t)a_{\alpha\beta}^{*}(t)\\
0&0&0&0
\end{pmatrix}
\label{a9}
\end{equation}

where:
\begin{equation}
 p_{\sigma_{x_{1}}}=\frac{1}{2}\bigl[|a_{\alpha\beta}(t)|^{2}+|a_{\alpha\alpha}(t)|^{2}\bigl]=\frac{1}{2}.
\label{a10}
\end{equation}

From(\ref{a9}) follows:

\begin{equation}
\rho_{B|\sigma_{x_{1}}}=
\begin{pmatrix}
|a_{\alpha\alpha}(t)|^{2}&a_{\alpha\alpha}(t)a_{\alpha\beta}^{*}(t)\\
a_{\alpha\beta}(t)a_{\alpha\alpha}^{*}(t)&|a_{\alpha\beta}(t)|^{2}
\end{pmatrix}
\label{a11}
\end{equation}

Remembering that:

\begin{equation}
\begin{split}
&\ket{y_{1}}=\frac{1}{\sqrt{2}}
\begin{pmatrix}
1\\
i
\end{pmatrix},
\hspace{0.3cm}
\ket{y_{2}}=\frac{1}{\sqrt{2}}
\begin{pmatrix}
1\\
-i
\end{pmatrix}\\
&\ket{z_{1}}=
\begin{pmatrix}
1\\
0
\end{pmatrix},
\hspace{0.5cm}
\ket{z_{2}}=
\begin{pmatrix}
0\\
1
\end{pmatrix}
\end{split}
\label{a12}
\end{equation}

it is simple to see that:

\begin{equation}
C^{\sigma_{x_{1}}}_{l_{1}}\bigl(\rho_{B|\sigma_{x_{k}}}\bigl)=1+4\sqrt{P_{\alpha\alpha}(t)(1-P_{\alpha\alpha}(t))}.
\label{a13}
\end{equation}

where $P_{\alpha\alpha}(t)=|a_{\alpha\alpha}(t)|^2$.

In the same way we calculate $C^{\sigma_{x_{2}}}_{l_{1}}\bigl(\rho_{B|\sigma_{x_{2}}}\bigl)$.\\
Likewise, the same procedure can be applied on performing Pauli measurement $\sigma_{y}$ or $\sigma_{z}$.

After all calculation, we find that:

\begin{equation}
N^{l_{1}}(\rho^{\alpha}_{AB}(t))=2+2\sqrt{(1-P_{\alpha\alpha}(t))P_{\alpha\alpha}(t)},
\label{a14}
\end{equation}

\subsection*{A.2: Bell nonlocality}
Now we see how to rewrite the Bell nonlocality criterion in terms of neutrino oscillation probability.

We calculate the  correlation matrix $T$ whose elements are $T_{m,n}=\Tr[\rho(\sigma_{m}\otimes \sigma_{n})]$, where $\sigma_{i}, i=1,2,3$ are the Pauli matrices.

\begin{equation}
T=
\begin{pmatrix}
a_{\alpha\beta}a_{\alpha\alpha}^{*}+a_{\alpha\alpha}a_{\alpha\beta}^{*}& -ia_{\alpha\beta}a_{\alpha\alpha}^{*}+ i a_{\alpha\alpha}a_{\alpha\beta}^{*}&0\\
ia_{\alpha\beta}a_{\alpha\alpha}^{*}- i a_{\alpha\alpha}a_{\alpha\beta}& a_{\alpha\beta}a_{\alpha\alpha}^{*}+a_{\alpha\alpha}a_{\alpha\beta}^{*}&0\\
0&0&-|a_{\alpha\alpha}|^{2}-|a_{\alpha\beta}|^{2}
\end{pmatrix}
\label{a15}
\end{equation}

It is simple to calculate:

\begin{equation}
T^{\dagger}=
\begin{pmatrix}
a_{\alpha\beta}a_{\alpha\alpha}^{*}+a_{\alpha\alpha}a_{\alpha\beta}^{*}& ia_{\alpha\beta}a_{\alpha\alpha}^{*}- i a_{\alpha\alpha}a_{\alpha\beta}^{*}&0\\
-ia_{\alpha\beta}a_{\alpha\alpha}^{*}+ i a_{\alpha\alpha}a_{\alpha\beta}& a_{\alpha\beta}a_{\alpha\alpha}^{*}+a_{\alpha\alpha}a_{\alpha\beta}^{*}&0\\
0&0&-|a_{\alpha\alpha}|^{2}-|a_{\alpha\beta}|^{2}
\end{pmatrix}
\label{a16}
\end{equation}

From the matrix product calculation $T^{\dagger}T$ we found that the eigenvalues of this matrix are: 
\begin{itemize}
\item[] $u_{1}=(-|a_{\alpha\alpha}|^{2}-|a_{\alpha\beta}|^{2})^{2}=(-P_{\alpha\alpha}-P_{\alpha\beta})^{2}=(-1)^{2}=1$,
\item[] $u_{2}=u_{3}=4a_{\alpha\alpha}a_{\alpha\beta}a_{\alpha\alpha}^{*}a_{\alpha\beta}^{*}=4P_{\alpha\alpha}(1-P_{\alpha\alpha})$,
\end{itemize}
where we have used $P_{\alpha\alpha}+P_{\alpha\beta}=1$.

\section*{Appendix B: Wave packet description of neutrino oscillations.}
\label{Appendice B}
In this appendix we briefly review the  wave packet approach to neutrino oscillations \cite{Giunti},\cite{Giunti2}.
\\
Let us consider a neutrino with definite flavor $\alpha (\alpha=e,\mu,\tau)$, that propagates along $x$ axis. We can write:

\begin{equation}
\ket{\nu_{\alpha}(x,t)}=\sum_{j}U_{\alpha j}^{*}\psi_{j}(x,t)\ket{\nu_{j}},
\label{sub1}
\end{equation}
where $U_{\alpha j}$ denotes the elements of the PMNS mixing matrix and  $\psi_{j}(x,t)$ is the wave function of the mass eigenstate $\ket{\nu_{j}}$ with mass $m_{j}$. If we assume a Gaussian distribution for the momentum of the massive neutrino ${\nu_{j}}$:

\begin{equation}
\psi_{j}(p)=\bigl(2\pi {\sigma_{p}^{P}}^{2}\bigl)^{-\frac{1}{4}}\exp{-\frac{(p-p_{j})^{2}}{4{\sigma_{p}^{P}}^{2}}}
\label{sub2}
\end{equation}
where $p_{j}$ is the average momentum and $\sigma_{p}^{P}$ is the momentum uncertainty determined by the production process, the wave function is:

\begin{equation}
\psi_{j}(x,t)=\frac{1}{\sqrt{2\pi}}\int dp \hspace{0.1cm}\psi_{j}(p)e^{ipx-iE_{j}(p)t}, 
\label{sub3}
\end{equation}
where the energy is $E_{j}(p)=\sqrt{p^{2}+m_{j}^{2}}$.
Now we assume that the Gaussian momentum  distribution (\ref{sub2}) is strongly peaked around $p_{j}$, that is, we assume the condition $\sigma_{p}^{P}\ll E_{j}^{2}(p_{j})/m_{j}$. This allows us to approximate the energy with:

\begin{equation}
E_{j}(p)\simeq E_{j} + v_{j}(p-p_{j}),
\label{sub4}
\end{equation}
where $ E_{j}=\sqrt{p_{j}^{2}+m_{j}^{2}}$ is the average energy and $v_{j}=\frac{\partial E_{j}(p)}{\partial p}\biggl|_{p=p_{j}}=\frac{p_{j}}{E_{j}}$ is the group velocity of the wave packet of the massive neutrino $\nu_{j}$.

Using these approximations we can perform an integration on $p$ of (\ref{sub3}), obtaining:

\begin{equation}
\psi_{j}(x,t)=\bigl(2\pi {\sigma_{x}^{P}}^{2}\bigl)^{-\frac{1}{4}} \exp \biggl[-iE_{j}t + ip_{j}x - \frac{(x-v_{j}t)^{2}}{4 {\sigma_{x}^{P}}^{2}}\biggl]
\label{sub5}
\end{equation}
where $\sigma_{x}^{P}=\frac{1}{2\sigma_{p}^{P}}$ is the spatial width of the wave packet.

At this point, by substituting (\ref{sub5}) in (\ref{sub1}) it is possible to obtain the density matrix operator by $\rho_{\alpha}(x,t)=\ket{\nu_{\alpha}(x,t)}\bra{\nu_{\alpha}(x,t)}$
which describes the neutrino oscillations in space and time. Although in laboratory experiments it is possible to measure neutrino oscillations in time through the measurement of both the production and detection processes, due to the long time exposure in time of the detectors  it is convenient to consider an average in time of the density matrix operator. In this way $\rho_{\alpha}(x)$ is the relevant density matrix operator and it can be obtained by a gaussian time integration

In the case of ultra-relativistic neutrinos, it is useful to consider the following approximations: $E_{j}\simeq E + \xi_{P}\frac{m_{j}^{2}}{2E}$, where $E$ is the neutrino energy in the limit of zero mass and $\xi_{P}$ is a dimensionless quantity that depends on the characteristics of the production process, $p_{j}\simeq E-(1-\xi_{P})\frac{m_{j}^{2}}{2E}$ and $v_{j}\simeq 1-\frac{m_{j}^{2}}{2E_{j}^{2}}$.
Considering these approximations,  $\rho_{\alpha}(x)$ becomes:

\begin{equation}
\rho_{\alpha}(x)=\sum_{j,k}U_{\alpha j}^{*}U_{\alpha k} \exp\biggl[-i \frac{\Delta m_{jk}^{2}x}{2E}-\biggl(\frac{\Delta m_{jk}^{2}x}{4 \sqrt{2} E^{2} \sigma_{x}^{P}}\biggl)^{2} - \biggl(\xi_{P}\frac{\Delta m_{jk}^{2}}{4 \sqrt{2} E \sigma_{p}^{P}}\biggl)^{2}\biggl]\ket{\nu_{j}}\bra{\nu_{k}},
\label{sub8}
\end{equation}
where $\Delta m_{jk}^{2}=m_{j}^{2}-m_{k}^{2}$.

Taking into account that the detection process take place at a distance $L$ from the origin of the coordinates,

 the transition probability is given by:

\begin{equation}
P_{\nu_{\alpha}\rightarrow \nu_{\beta}}(L)= \Tr\bigl(\rho_{\alpha}(x)\mathcal{O}_{\beta}(x-L)\bigl)
= \sum_{j,k}U_{\alpha j}^{*}U_{\alpha k}U_{\beta j}^{*}U_{\beta k}\exp\biggl[-2\pi i \frac{L}{L^{osc}_{jk}}-\biggl(\frac{L}{L^{coh}_{jk}}\biggl)^{2}- 2 \pi^{2}\xi^{2}\biggl(\frac{\sigma_{x}}{L^{osc}_{jk}}\biggl)^{2}\biggl],
\label{sub10}
\end{equation}

where $L^{osc}_{jk}$ is the oscillation length and $L^{coh}_{jk}$  the coherence length, defined by:

\begin{equation}
L^{osc}_{jk}=\frac{4 \pi E}{\Delta m_{jk}^{2}}, \hspace{1cm} L^{coh}_{jk}=\frac{4 \sqrt{2} E^{2}}{|\Delta m_{jk}^{2}|}\sigma_{x},
\label{sub11}
\end{equation}
with $\sigma_{x}^{2}={\sigma_{x}^{P}}^{2} + {\sigma_{x}^{D}}^{2}$ and $\xi^{2}\sigma_{x}^{2}=\xi_{P}^{2}{\sigma_{x}^{P}}^{2} + \xi_{D}^{2}{\sigma_{x}^{D}}^{2}$,where $\sigma^{D}$ is the uncertainty of the detection process and $\xi_{D}$  depends from the characteristics of the detection process. \\

We note that the  wave packet description confirms the standard value of the oscillation length.  The coherence length is the distance beyond which the interference of the massive neutrinos $\nu_{j}$ and $\nu_{k}$ is suppressed. This because the separation of their wave packets when they arrive at the detector is so large that they cannot be absorbed coherently. The last term in the exponential of (\ref{sub11}) implies that the interference of the neutrinos is observable only if the localization of the production and detection processes is smaller than the oscillation length.\\

\end{document}